\documentclass[final, 1p, times,nopreprintline]{elsarticle}

\usepackage{amssymb}

\usepackage{lineno}

\usepackage{array,multirow}
\usepackage{subcaption}

\usepackage{xcolor}
\usepackage[normalem]{ulem}

\usepackage{siunitx}

\usepackage{todonotes}

\newcommand{\Li}{$^{8}$Li~}

\newcommand{\B}{$^{8}$B~}

\journal{Nuclear Instruments and Methods A}

\begin{document}

\begin{frontmatter}



\title{The Beta-decay Paul Trap Mk IV: design and commissioning}


\author[uchicago,anl]{L. Varriano\corref{cor1}}
\ead{varriano@uchicago.edu}
\cortext[cor1]{Corresponding author present address: University of Washington, Center for Experimental Nuclear Physics and Astrophysics, Box 354290, Seattle, WA, 98195, USA. Tel:+1 206 543 4080 }

\author[uchicago,anl]{G. Savard} 
\author[anl,manitoba]{J. A. Clark} 
\author[anl]{D. P. Burdette} 
\author[llnl]{M. T. Burkey} 
\author[llnl]{A. T. Gallant} 
\author[soreq]{T. Y. Hirsh} 
\author[llnl]{B. Longfellow} 
\author[llnl]{N. D. Scielzo} 
\author[northwestern]{R. Segel}

\author[anleof]{E. J. Boron, III} 
\author[nd]{M. Brodeur} 
\author[anl]{N. Callahan} 
\author[nd]{A. Cannon} 
\author[llnl]{K. Kolos} 
\author[nd]{B. Liu} 
\author[lsu]{S. Lopez-Caceres} 
\author[anl]{M. Gott\fnref{mattornl}} 
\author[anl]{B. Maa{\ss}} 
\author[lsu]{S. T. Marley}
\author[anl]{C. Mohs} 
\author[lsu]{G. E. Morgan} 
\author[anl]{P. Mueller}
\author[anl]{M. Oberling} 
\author[nd]{P. D. O'Malley} 
\author[nd]{W. S. Porter} 
\author[lsu]{Z. Purcell} 
\author[anl,manitoba]{D. Ray\fnref{raytriumf}} 
\author[nd]{F. Rivero} 
\author[anl,manitoba]{A. A. Valverde} 
\author[lsu]{G. L. Wilson\fnref{gemmanow}}
\author[nd]{R. Zite} 

\affiliation[uchicago]{
            organization={Department of Physics, University of Chicago},
            city={Chicago},
            state={IL},
               postcode={60637}, 
            country={USA}}
            
\affiliation[anl]{
            organization={Physics Division, Argonne National Laboratory},
            city={Lemont},
            state={IL},
              postcode={60439}, 
            country={USA}}

\affiliation[manitoba]{
            organization={Department of Physics and Astronomy, University of Manitoba}, 
            city={Winnipeg}, 
            state={Manitoba}, 
            postcode={R3T 2N2}, 
            country={Canada}}
            
\affiliation[llnl]{
            organization={Lawrence Livermore National Laboratory},
            city={Livermore}, 
            state={CA}, 
            postcode={94550}, 
            country={USA}}

\affiliation[soreq]{
            organization={Soreq Nuclear Research Center}, 
            state={Yavne}, 
            postcode={81800}, 
            country={Israel}}

\affiliation[northwestern]{
            organization={Department of Physics and Astronomy, Northwestern University}, city={Evanston}, 
            state={IL}, 
            postcode={60208}, 
            country={USA}}

\affiliation[anleof]{
            organization={Experimental Operations and Facilities, Argonne National Laboratory},
            city={Lemont},
            state={IL},
              postcode={60439}, 
            country={USA}}
            
\affiliation[nd]{
            organization={Department of Physics and Astronomy, University of Notre Dame}, 
            city={Notre Dame}, 
            state={IN}, 
            postcode={46556}, 
            country={USA}}
     
\affiliation[lsu]{
            organization={Department of Physics and Astronomy, Louisiana State University}, 
            city={Baton Rouge},
            state={LA}, 
            postcode={70803}, 
            country={USA}}

\fntext[mattornl]{Present Address: Enrichment Science and Engineering Division, Oak Ridge National Laboratory, Oak Ridge, TN, 37830, USA}
\fntext[raytriumf]{Present Address: Physics Department, McGill University, Montreal, Quebec H3A 2T8, Canada and TRIUMF, Vancouver, British Columbia V6T 2A3, Canada}
\fntext[gemmanow]{Present Address: United Kingdom Atomic Energy Authority, Culham Centre for Fusion Energy, Culham Science Centre, Abingdon, Oxon OX14 3DB, United Kingdom}

\begin{abstract}
The Beta-decay Paul Trap is an open-geometry, linear trap used to measure the decays of \Li and \B to search for a tensor contribution to the weak interaction. In the latest \Li measurement of Burkey et al. (2022) \cite{Burkey_Savard_Gallant_Scielzo_Clark_Hirsh_Varriano_Sargsyan_Launey_Brodeur_etal._2022}, $\beta$ scattering was the dominant experimental systematic uncertainty.  The Beta-decay Paul Trap Mk IV reduces the prevalence of $\beta$ scattering by a factor of 4 through a redesigned electrode geometry and the use of glassy carbon and graphite as electrode materials. The trap has been constructed and successfully commissioned with \Li in a new data campaign that collected $2.6$ million triple coincidence events, an increase in statistics by 30\% with 4 times less $\beta$ scattering compared to the previous \Li data set.
\end{abstract}




\end{frontmatter}


\section{Introduction}
\label{sec:sample1}


Ion traps are an attractive tool to study short-lived isotopes of interest to nuclear physics and have been used successfully for decades (for a recent review, see \cite{Eronen_Kankainen_Äystö_2016}). These devices offer an advantage over implantation techniques for $\beta$-decay correlation measurements since the nucleus decays in free space and all decay products (sans $\nu$) can be detected, including the recoiling ion. In contrast to atom traps, any element can be trapped in the same instrument, in principle, and the trapping efficiency can be as high as 100\%, provided that the half-life of the ion of interest is sufficiently long for cooling, transport, and trapping, typically $\gtrsim$ 10 ms. Additionally, no energy is lost to the surrounding medium as in implantation techniques, allowing for energy and momentum reconstruction of the decay products. Several different recent ion trap experiments \cite{Fléchard_2011, SCIELZO201470} have detected both the $\beta^{\pm}$ from the initial decay and the recoiling ion, a technique that will be used in future ion traps, as well \cite{mora, tamutrap, stbenedict, beartrap, eibt}.

The Beta-decay Paul Trap (BPT) is a linear Paul trap at the Argonne Tandem Linac Accelerator System (ATLAS) at Argonne National Laboratory that has been designed for precision tests of the Standard Model (SM) \cite{Scielzo_2012, Burkey_2019} and has also been used to study $\beta$-delayed neutron emission \cite{wangbpt, PhysRevLett.110.092501, PhysRevC.101.024312, Siegl_2018}. The BPT is used to study the decays of the mirror nuclei $^8$Li and $^8$B to search for a tensor contribution to the weak interaction \cite{Burkey_Savard_Gallant_Scielzo_Clark_Hirsh_Varriano_Sargsyan_Launey_Brodeur_etal._2022, Li_2013, Sternberg_2015, 8b}. Such a contribution arises in various beyond-Standard Model (BSM) extensions \cite{Herczeg_2001} and appears at the quark level through the Wilson coefficients $C_X^{(\prime)}$ that describe the coupling strengths for the several possible Lorentz-invariant forms: $X = V, A, S, T, P$, which are vector, axial-vector, scalar, tensor, and pseudoscalar, respectively. Both \Li and \B are essentially pure Gamow-Teller decays \cite{Wiringa_Pastore_Pieper_Miller_2013} that proceed through an axial-vector $(A)$ current in the SM and a possible tensor $(T)$ current . Combining the results of the \Li and \B mirror decays also allows for a joint constraint on the Fierz interference term $b_{\mathrm{Fierz}}$ \cite{8b}. For a recent comprehensive analysis of BSM searches with nuclear decays, see Ref. \cite{Falkowski_González-Alonso_Naviliat-Cuncic_2021}. The $^8$B neutrino spectrum can also be reconstructed, which is important for the interpretation of solar neutrino astrophysics experiments \cite{brenden}. 

Both $^8$Li (J$^\pi$ = 2$^{+}$, isospin T = 1, Q$_{\beta}$=16.00413(6) MeV) and $^8$B (2$^{+}$, 1, 16.9579(10) MeV) predominantly decay to a broad 3 MeV excited state in $^8$Be (2$^{+}$, 0), which $\alpha$-decays within $\sim10^{-22}$ seconds \cite{AME2020}. This allows the experiment to detect an essentially background-free triple correlation of the 2 $\alpha$ particles and the $\beta$ from the initial decay. The short $\alpha$-decay half-life and high $\alpha$ energy ensures that the momentum of the decay products is essentially undisturbed by the trapping potential of a few hundred volts. Detecting the energy and momenta of the two $\alpha$ and the direction of the $\beta$ allows for a kinematically-complete reconstruction of the decay; four double-sided silicon strip detectors are used in the BPT and described in more detail below. The high Q$_{\beta}$ value and light nuclear mass also means that the $\alpha$ particles can have energy differences of up to $\sim400$ keV and can have momenta offset from anti-parallel by up to $\sim 20^{\circ}$ in the laboratory frame. In practice, a measurement is performed by comparing the $\alpha$ energy difference spectrum to simulations. In the measurement of the $\alpha$ energy difference spectrum, the BPT exploits a triple correlation enhancement by selecting only events in which the $\beta$ was roughly parallel to an $\alpha$. This decay geometry results in larger average recoil energies under a $T$ interaction than an $A$ interaction, due to the alignment of the lepton momenta \cite{Sternberg_2015}. In addition, taking the difference between $\alpha$ energies reduces systematic uncertainties associated with an imperfect knowledge of the detector response. By subtracting the energy of the lone $\alpha$ from the energy of the $\alpha$ roughly parallel to the $\beta$, uncertainties in the detector response are reduced, including both those that are common to all detectors as well as individual detector uncertainties.

The ultimate goal of BPT is to measure $|C_T / C_A|^2$ to an uncertainty of $1 \times 10^{-3}$ or better under an assumption of a right-handed $T$ interaction. The most recently published \Li measurement with the BPT obtained $|C_T / C_A|^2 = 0.0012 \pm 0.0019_{\mathrm{stat}} \pm 0.0028_{\mathrm{syst}}$, a result consistent with the SM and the most stringent low-energy limit on a tensor contribution to date \cite{Burkey_Savard_Gallant_Scielzo_Clark_Hirsh_Varriano_Sargsyan_Launey_Brodeur_etal._2022}. The subsequent break-up of the recoiling $^8$Be$^*$ into 2 $\alpha$ particles is described in detail in Ref. \cite{Holstein_1974}, which is the formulation used in the simulation and analysis of the BPT experiments. The uncertainty from recoil-order terms has recently been improved by a factor of 2 through a new calculation of the recoil-order parameters \cite{Sargsyan2022}. Radiative corrections for the decay are described in Ref. \cite{Gluck_1997}. Further theoretical improvements are expected, and an experimental measurement of recoil-order parameters may further reduce the uncertainty on these terms

Among the experimental systematic uncertainties, the largest source comes from $\beta$ scattering, which is also an important consideration in similar experiments \cite{Fléchard2011}. The analysis of the data collected with the BPT focuses on events where a $\beta$ hits the same detector as an $\alpha$. However, if a $\beta$ scatters off of the trap structure into a detector coincident with an $\alpha$ pair, the decay kinematics will be incorrectly reconstructed. With the current BPT detector set-up, it is not possible to remove scattered-$\beta$ events through cuts on the kinematic reconstruction, as a physically-allowed combination of momenta can nearly always be obtained. This effect is modeled with a detailed \textsc{Geant4} simulation using the ``option3” standard electromagnetic physics list \cite{AGOSTINELLI2003250,1610988,ALLISON2016186}. In the latest \Li experiment \cite{Burkey_Savard_Gallant_Scielzo_Clark_Hirsh_Varriano_Sargsyan_Launey_Brodeur_etal._2022}, a significant fraction of events, roughly 21\%, were from scattered $\beta$s.  Due to the kinematically-complete events detected in the BPT, it would be possible to directly measure some or all of the recoil-order parameters by analyzing other spectra that are sensitive to these terms, such as the angular distribution between the $\alpha$ and the $\beta$. However, events from $\beta$ scattering broaden these spectra, and thus previous experiments with the BPT do not have sufficient precision to reduce the uncertainties on these parameters. 

To reach the measurement goal of $1 \times 10^{-3}$ uncertainty on $|C_T / C_A|^2$, $\beta$ scattering needs to be dramatically reduced. This reduction has been achieved with the Beta-decay Paul Trap Mk IV, a newly-designed linear Paul Trap that utilizes carbon electrodes and a new geometry to reduce the effect of $\beta$ scattering by a factor of $\sim$4. Coupled with a parallel effort to improve the detector characterization with $\alpha$ beams, the BPT Mk IV aims to reduce experimental systematic uncertainties by a factor of $\sim$2 to enable further improvement to the precision of the measurement.

\section{BPT description} \label{sec:trapbasics}

The BPT has an open-geometry and uses a quadrupole arrangement of electrodes with voltages oscillating at radio-frequencies (RF) to confine ions in the radial plane and uses segmented, static voltage (DC) electrodes to confine ions in the axial direction. An ideal Paul trap uses hyperbolic electrodes to maximize the physical extent of the trapping potential while minimizing its anharmonicity. Any arrangement of electrodes with quadrupole symmetry will have a hyperbolic potential near its center, enabling the electrode shape to have a more open geometry to allow for optical access to the ion cloud. Different electrode geometries, however, will have an impact on the size of this hyperbolic region, as discussed in section \ref{sec:redesign}. Previous changes to the trap electrodes and experimental components have been noted in Refs. \cite{Burkey_2019, Sternberg_2013}.

The entire apparatus meets ultra-high vacuum (UHV) standards to maintain an environment free from out-gassing that might affect the trapping lifetime. The typical pressure achieved prior to cryogenic cooling is $\lesssim 1\times10^{-8}$ mbar. The BPT uses ultra-pure 99.999\% helium gas at a $\sim 10^{-5}$ mbar pressure to cool ions through collisions with the gas. The frame of the trap is hollow and liquid nitrogen is circulated to bring the trap and gas to approximately 80 K \cite{ln2pump}: in this way, the ions of interest are thermalized to this lower temperature, reducing the overall size of the ion cloud. A low thermalization temperature is critical for the experiment; from the Maxwell-Boltzmann distribution at 80 K, 99\% of ions have an energy less than $\sim40$ meV, compared to $\sim150$ meV at room temperature. Therefore, the voltage required to contain the ions in the same volume is almost a factor $2$ smaller at this lower temperature, reducing RF pickup on the detectors. In addition, the cryogenic temperature greatly reduces out-gassing of trap construction materials, leading to a longer trap lifetime of at least tens of seconds \cite{Scielzo_2012}. 

The ion capture, transport, and delivery system are described in Ref. \cite{Scielzo_2012}. The BPT operates by accumulating ion bunches to build up an equilibrium population during data-taking before ejecting all of the ions to perform a background measurement. The typical cycle for \Li is comprised of 70 ion bunch injections, one every 160 ms, followed by an 800 ms background measurement, for a total of a 12 second measurement cycle. Data is ignored during the first $\sim 30$ ms following each injection, during which the ion cloud thermalizes with the buffer gas, to avoid systematic effects associated with a larger ion cloud.

Four Micron BB7 double-sided silicon strip detectors (DSSDs) \cite{dssd} surround the trap at a distance of 6.5 cm from the center, with a total of 25\% solid angle coverage. These 32x32 strip detectors have a pitch of 2 mm with a nominal 100 nm aluminized dead layer covering most of the strip with a thicker, more complicated structure around the 25 \si{\micro\meter} interstrip gaps. The DSSD thickness is 1 mm. The cryogenic cooling provides for lower leakage currents, giving improved resolution. To mitigate RF pickup on the DSSDs, tunable notch filters are included before preamplification and long shaping times are used  \cite{Burkey_2019}. The DSSDs have a typical $\alpha$-particle resolution of 20-30 keV FWHM. The minimum-ionizing $\beta$ particles deposit a few hundred keV in the DSSD, and a plastic scintillator detector behind each DSSD detects the remaining $\beta$ energy. 

\section{$\beta$ scattering in previous BPT \label{sec:scattering}}

The BPT has an open geometry, but material and design choices still strongly impact $\beta$ scattering. A modified version of the BPT \textsc{Geant4} (release 10.5.1) simulation code was used to study the effect of different design choices on $\beta$ scattering. A CAD model of the BPT was designed in Autodesk Inventor \cite{autodesk} then exported to STP files. Using a modified version of an existing script \cite{cad-to-Geant4-converter}, these files are then converted to GDML files readable by \textsc{Geant4}. An ion cloud of realistic dimensions (obtained from previous experiments) and decay spectrum was simulated. Events where both $\alpha$ particles hit opposite DSSDs and the $\beta$ hits a coincident detector were used to determine the scattered ratio (triple coincidence events prior to final analysis cuts). If a detector hit of a $\beta$ did not match its true original momentum, it was counted as scattered. Statistical uncertainty on the scattered fractions reported below is $\sim$0.1\% with $\sim$150,000 triple coincidences from each simulation.

The previous BPT was found to have 21.1\% of triple coincidence events from scattered $\beta$ particles. It is not entirely possible to cleanly separate physical sources of scattering, but with the \textsc{Geant4} model, the major contributions can be understood. Of the total 21.1\% scattering, 5.5\% is attributed to the 0.01" thick stainless steel ``RF outer shields" (Fig. \ref{fig:oldview3}) used in an attempt to reduce to RF pickup on the DSSDs \cite{Burkey_2019, Sternberg_2013}; these outer shields were later found to have minimal impact on RF pickup reduction and therefore were removed for the BPT Mk IV design. Additional metal elements surrounding the DSSDs (0.006" thick stainless steel ``RF hoods" and 1/16" thick gold-plated aluminum detector covers, Fig. \ref{fig:oldview3}) were found to have some RF shielding efficacy but contributed an additional 5.5\% to $\beta$ scattering. These elements were redesigned. A further 5.6\% is attributed to the 2 mm thick stainless steel flat electrodes (Fig. \ref{fig:oldview2}), primarily the center electrode that forms the axial potential well (Fig. \ref{fig:oldview1}). The redesign of the electrodes is the focus of the next section. The remaining 4.5\% comes from back-scattering on the DSSDs and scattering on the electrode support structure. This last portion is largely irreducible without an extensive re-design of the vacuum vessel.


\section{Design of the BPT Mk IV} \label{sec:redesign}

\begin{figure}
     \centering
     \hfill
     \begin{subfigure}[b]{0.49\textwidth}
         \centering
         \includegraphics[page=2,trim={0cm 3cm 0cm 3cm},width=\textwidth,clip]{trapcomparison.pdf}
       \caption{previous BPT electrodes (beam axis horizontal)}
         \label{fig:oldview1}
     \end{subfigure}
     \hfill
     \begin{subfigure}[b]{0.49\textwidth}
         \centering
         \includegraphics[page=1,trim={0cm 3cm 0cm 3cm},width=\textwidth,clip]{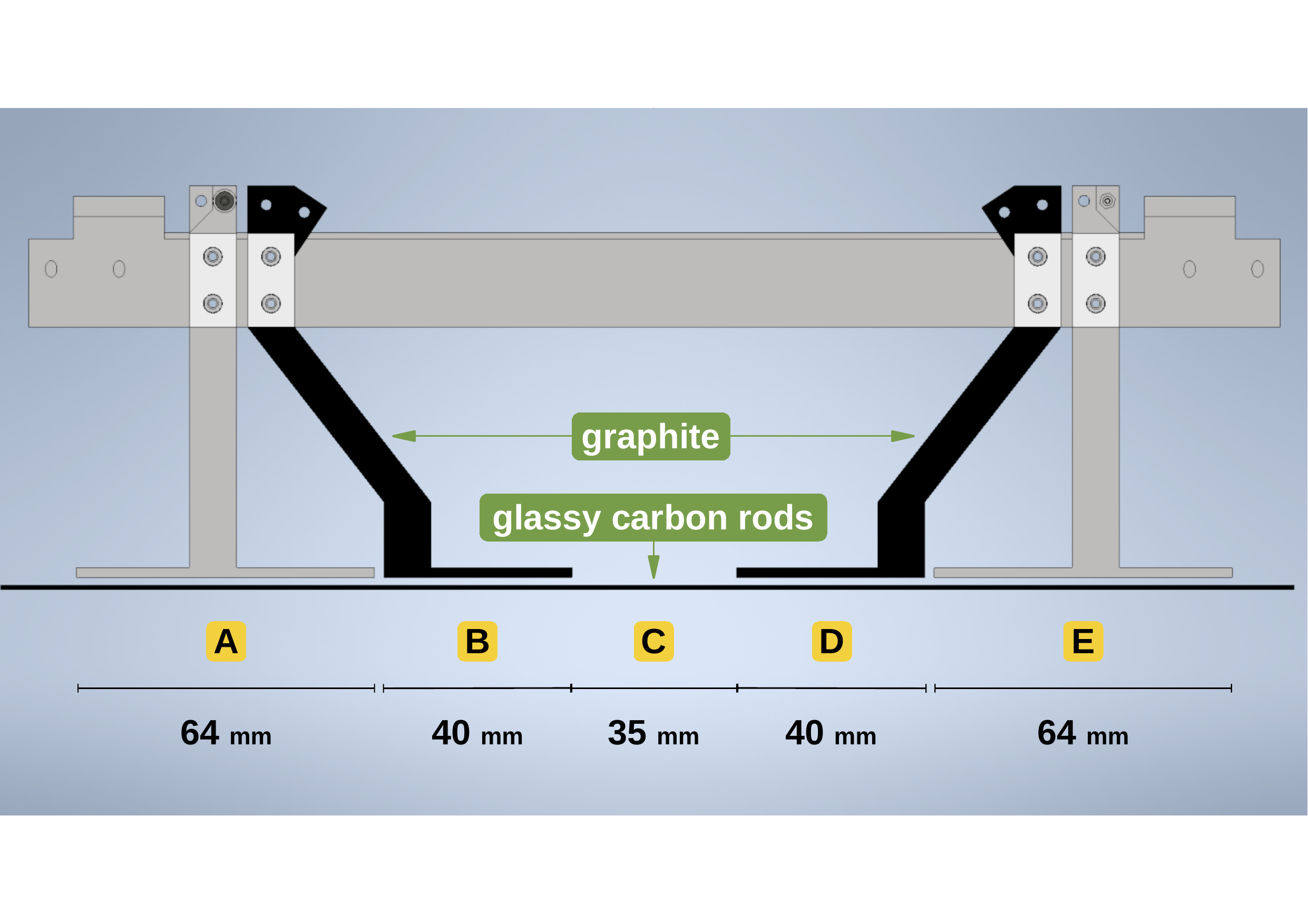}
       \caption{BPT Mk IV electrodes (beam axis horizontal) }
         \label{fig:mkivview1}
     \end{subfigure}
     \hfill

     \vspace{0.25cm}

     \hfill
     \begin{subfigure}[b]{0.49\textwidth}
         \centering
         \includegraphics[page=3,trim={0cm 3cm 0cm 3cm},width=\textwidth,clip]{trapcomparison.pdf}
       \caption{previous BPT electrodes (along beam axis)}
         \label{fig:oldview2}
     \end{subfigure}
     \hfill
     \begin{subfigure}[b]{0.49\textwidth}
         \centering
         \includegraphics[page=4,trim={0cm 3cm 0cm 3cm},width=\textwidth,clip]{trapcomparison.pdf}
       \caption{BPT Mk IV electrodes (along beam axis)}
         \label{fig:mkivview2}
     \end{subfigure}
     \hfill

          \vspace{0.25cm}

     \hfill
     \begin{subfigure}[b]{0.49\textwidth}
         \centering
         \includegraphics[page=5,trim={0cm 3cm 0cm 3cm},width=\textwidth,clip]{trapcomparison.pdf}
       \caption{previous BPT detector assembly}
         \label{fig:oldview3}
     \end{subfigure}
     \hfill
     \begin{subfigure}[b]{0.49\textwidth}
         \centering
         \includegraphics[page=6,trim={0cm 3cm 0cm 3cm},width=\textwidth,clip]{trapcomparison.pdf}
       \caption{BPT Mk IV detector assembly}
         \label{fig:mkivview3}
     \end{subfigure}
     \hfill
     
	\caption{Comparison of major design elements between the previous BPT, as used in Refs. \cite{Burkey_Savard_Gallant_Scielzo_Clark_Hirsh_Varriano_Sargsyan_Launey_Brodeur_etal._2022, brenden}, and the BPT Mk IV, performed in Autodesk Inventor \cite{autodesk}. Unlabeled components are made of stainless steel (grey) or alumina ceramic (white). (a) and (b) compare a single quadrant of the electrodes with the beam axis horizontal. The different DC voltage regions in the Mk IV are lettered in yellow; the ion cloud sits in the center region. (c) and (d) compare the electrodes along the beam axis; note that the cross section in the BPT Mk IV changes along the beam axis---cf. (b)---and only the rods are present in the center trapping region. The voltage pattern applied to the Mk IV is labeled in yellow. (e) and (f) compare a sectioned view of the detector and RF shielding assemblies. See text for additional details.}
	\label{fig:comparetraps}
\end{figure}

\begin{figure}
\centering
\includegraphics[page=7,trim={4cm 0cm 4cm 0cm},width=0.8\textwidth,clip]{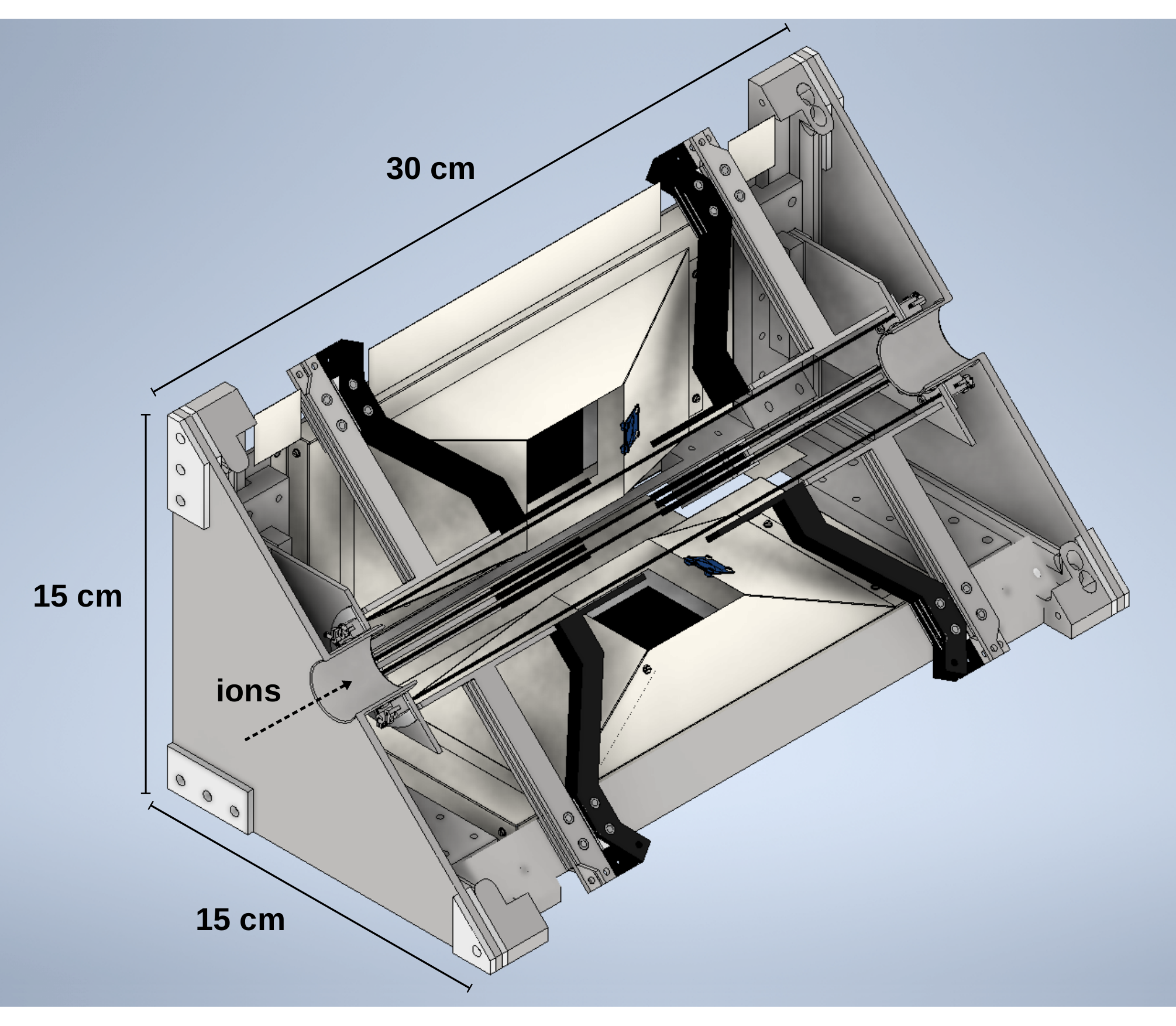}
\caption{Sectioned view of the final design of the BPT Mk IV performed in Autodesk Inventor \cite{autodesk}. See Fig. \ref{fig:comparetraps} and text for additional details. }
\label{fig:trap} 
 \end{figure}

 The design of the BPT Mk IV is shown in Fig. \ref{fig:trap}.  A detailed comparison of major design elements between the previous BPT and the BPT Mk IV is shown in Fig. \ref{fig:comparetraps}. The design goals for the BPT Mk IV were the reduction of both $\beta$ scattering and RF pickup on the DSSDs \cite{Burkey_2019}. These two goals compete with each other since a reduction in scattering necessitates removing material close to the ion cloud, while reducing the necessary RF amplitude requires moving the electrodes closer to the ion clouds and/or adding material surrounding the DSSDs to reduce RF pickup. The new design also includes five electrode segments along the trap axis (``DC electrode regions") in an effort to reduce disturbing the ion cloud during capture pulses. The previous BPT trap design used three DC electrode regions, leading to some heating of the ion cloud because the potential at the trap center was perturbed during each capture pulse. 

The other major redesigned elements are the detector covers and RF hoods (Fig. \ref{fig:mkivview3}), which provide RF shielding of the DSSDs. These are made of 0.002" full-hard stainless steel, which was the thinnest sheet metal found to hold a rigid shape. The RF hood also has a much shallower angle that minimizes the effective thickness of metal presented to $\beta$ particles coming from the trap center. Compared to the previous electrodes and detector assembly (Fig. \ref{fig:comparetraps}), this design has much less material near the trapped ion cloud, reducing the chance of $\beta$ scattering. In the previous BPT, the calibration $\alpha$ sources faced detectors across the trapping volume, and two sources each illuminated half of the DSSD with significant overlap in the middle of the detector. Additionally, a quarter of the detector was not illuminated, making calibration difficult on these strips. In the new design, a single calibration $\alpha$ source illuminated an entire DSSD at a better-known position. As part of the overall effort to minimize the amount of material in the vicinity of the ion trap, the stainless steel frame rails that support the electrodes and the detector mounts were also made thinner. The endplates have a cylindrical hole and are held at a low DC potential to allow the ions to make the transition from the drift tube of the beamline to the trap (Fig. \ref{fig:trap}). 

The RF electrodes were designed so that they did not block the line of sight between the ion cloud and the DSSDs. The ion cloud was assumed to have maximal dimensions of 1.5mm × 1.5mm × 1.5mm, a realistic size from previous traps. Tolerance to allow for a slightly larger size was included in the final design.  SIMION v8.1 was used to model the electric potential for different electrode designs \cite{simion}. A grid spacing of 0.05 mm/grid unit was required to accurately determine the potential differences between rods of $\lesssim 1$ mm diameter. The basic electrode design premise was that the center electrode region, where the ion cloud is held, labeled region ``C" in Fig. \ref{fig:mkivview1}, should be formed by wires or rods supported at the ends of the trap. This arrangement entirely removes the support structure near the critical scattering region nearest to the DSSDs (Fig. \ref{fig:trap}). As shown in Fig. \ref{fig:mkivview2}, the rods supply the RF potential radially; the DC electrodes provide trapping along the axis. It was found that the DC electrode structure needed to ``reinforce" the RF radial potential, hence the pattern of applied voltages shown in Fig. \ref{fig:mkivview2}.

\begin{figure}
\includegraphics[width=\linewidth,trim={2cm 0 2cm 2cm}]{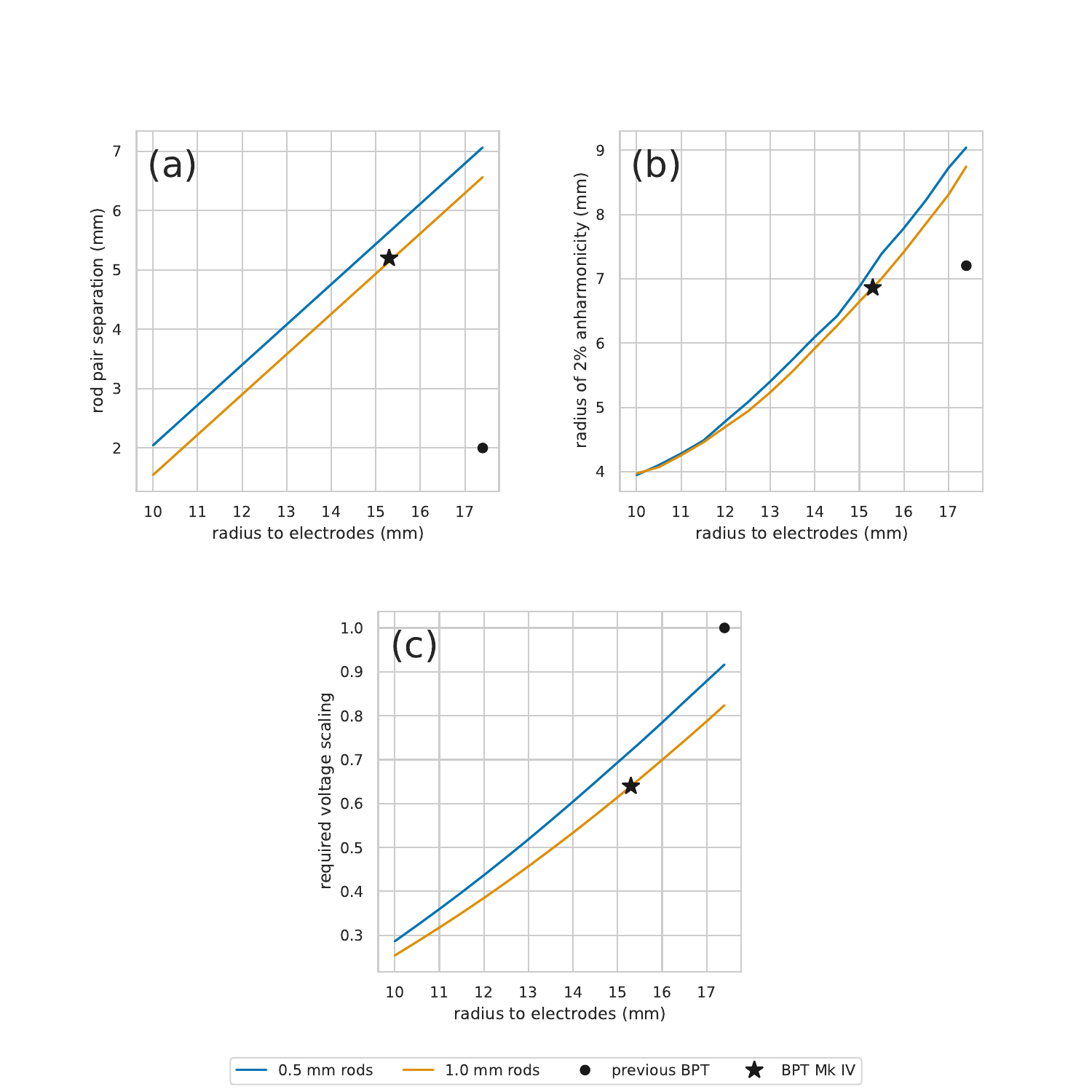}
\caption{ Comparison between 0.5 mm rod (blue) and 1.0 mm rod (orange) designs; (a) separation between pairs of electrode rods as calculated from the allowed electrode region; (b) radius at which terms higher than quadratic contribute $\geq$2\% to the potential; (c) required voltage relative to previous BPT to achieve same radial potential.}
\label{fig:quad}
 \end{figure}

To better approximate the curvature of an ideal hyperbolic electrode shape, tests were conducted with 2 rods spaced as widely as possible in the allowable electrode region (Fig. \ref{fig:quad}a). A 3 rod design spaced in a more hyperbolic shape was also tested. No significant difference in the electric potential near the trap center was found between the 2 and 3 rod designs, and scattering favored as few structures as possible. Thin metal wires, such as gold-plated tungsten wires common in time projection chambers, were considered, but due to the engineering challenges involved, a simpler design using larger diameter ($\sim 1$ mm), more rigid rods was favored. Following a similar approach as in Ref. \cite{mora_2020}, the potential calculated along the radial direction from the center to the electrodes was fit to even polynomial terms of order $r^0$ to $r^6$. The radius at which terms higher than quadratic order contribute more than 2\% to the total potential is then shown in Fig. \ref{fig:quad}b. Anharmonicities from these higher order contributions lead to greater instability in the ion motion, leading to fewer trapped ions \cite{Delahaye_Ban_Benali_Durand_Fabian_Fléchard_Herbane_Liénard_Mauger_Méry_etal._2019}. This treatment provides a metric for determining the stable trapping region for capture before the ions cool to a much smaller radius. The final design selected rods of 1.0 mm diameter at a distance of 15.5 mm from the center with a separation of 5.2 mm (Fig. \ref{fig:mkivview2}), which achieves a similar ion acceptance region as the previous BPT while requiring only 68\% of the previous BPT RF amplitude (Fig. \ref{fig:quad}c).

\begin{figure}
\centering

     \hfill
     
     \begin{subfigure}[b]{0.6\textwidth}
         \centering
         \includegraphics[trim={0cm 0cm 0cm 0cm},width=\textwidth,clip]{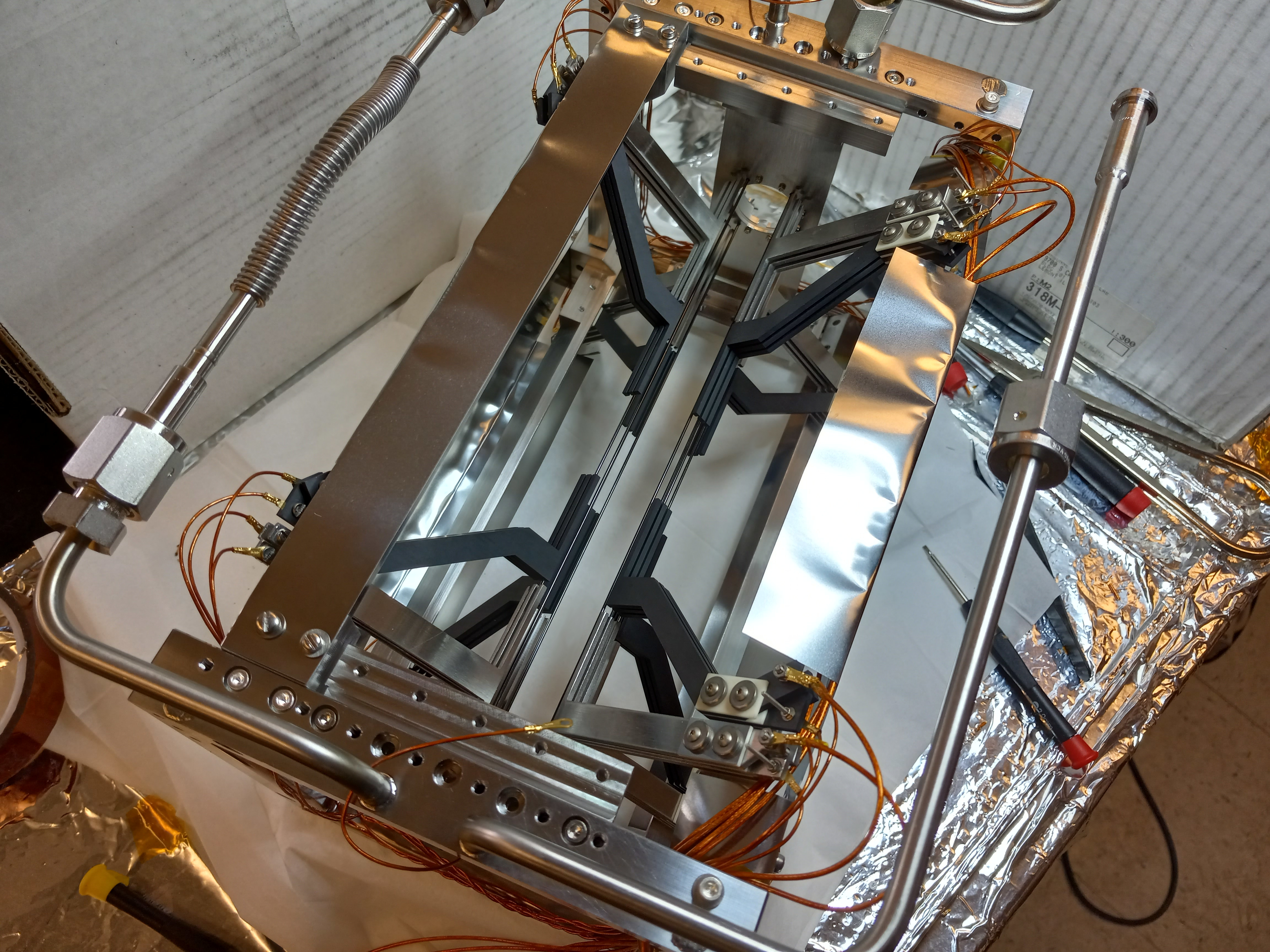}
       \caption{BPT Mk IV during assembly}
         \label{fig:mkivphysical1}
     \end{subfigure}
     
     \hfill

          \vspace{0.25cm}
          
     \hfill
     \begin{subfigure}[b]{0.49\textwidth}
         \centering
         \includegraphics[trim={0cm 0cm 0cm 0cm},width=\textwidth,clip]{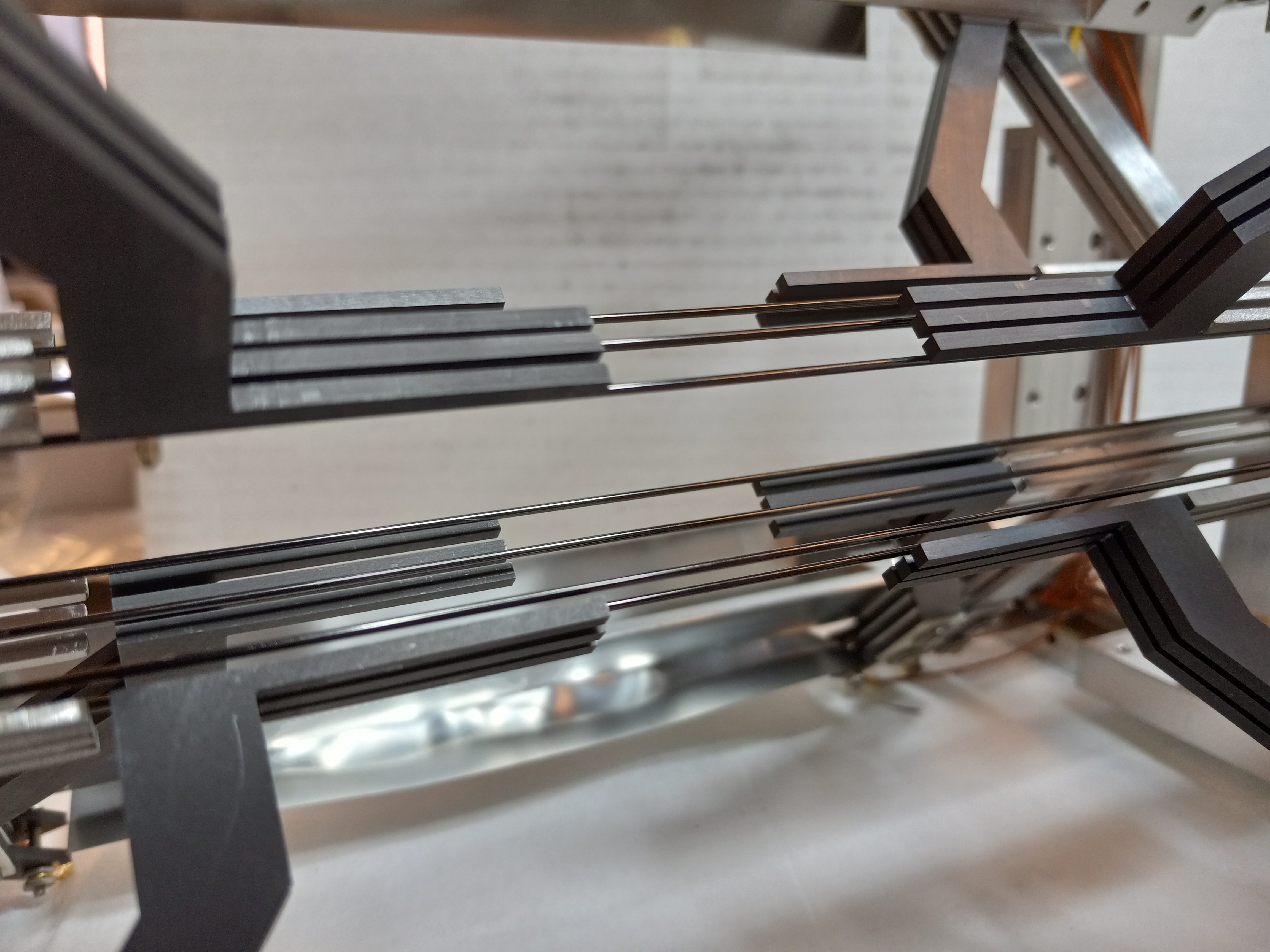}
       \caption{graphite and glassy carbon electrodes}
         \label{fig:mkivphysical2}
     \end{subfigure}
     \hfill
     \begin{subfigure}[b]{0.49\textwidth}
         \centering
         \includegraphics[trim={0cm 0cm 0cm 0cm},width=\textwidth,clip]{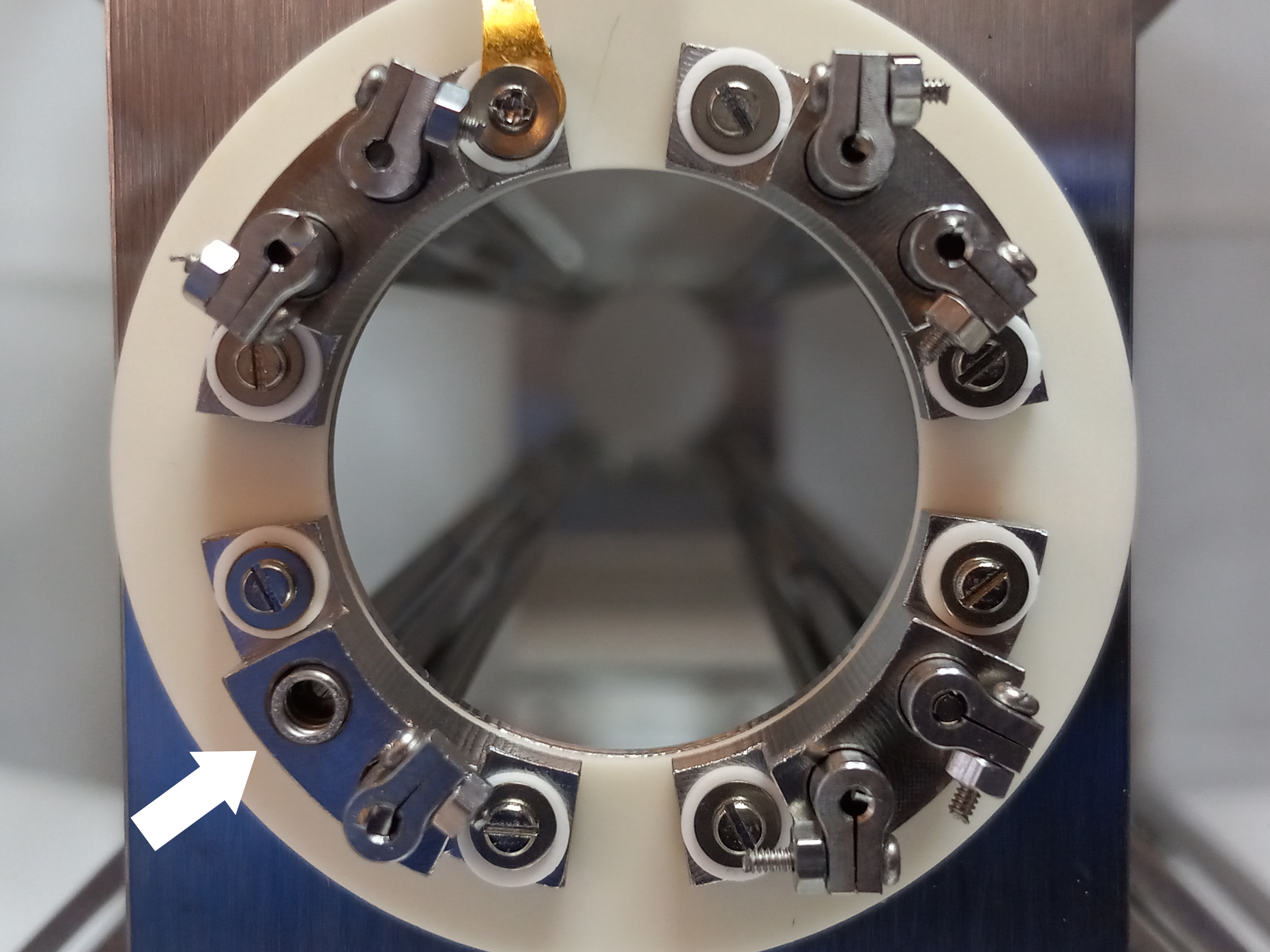}
       \caption{glassy carbon rod support structure}
         \label{fig:mkivphysical3}
     \end{subfigure}
     \hfill

\caption{The BPT Mk IV during assembly. (a) Full view of the trap with liquid nitrogen circulation lines shown. (b) A view of the center trapping region. (c) Support structure for the glassy carbon rods, which are held by friction in a small collar. A spring, indicated by a white arrow in lower left, is inserted into each alignment hole to tension the rods. }
\label{fig:bpt}
 \end{figure}

In the energy range of interest ($< 17$ MeV), the cross section for $\beta$ scattering increases with target material proton number as $Z^2$, so a low $Z$ material is intrinsically desirable \cite{Koch_Motz_1959}. Glassy carbon rods are commercially available with diameters as small as 1.0 mm and were found to contribute a relatively low amount of scattering due to both their low $Z$ and relatively low density of 1.42 g/cm$^3$ \cite{htw}. Glassy carbon is non-porous, chemically inert, and electrically conductive, making it an ideal material.  A key design challenge was the support structure for the glassy carbon rods, as these rods are flexible yet easy to shatter. In addition, the thermal contraction of the stainless steel frame amounts to about 0.8 mm reduction in length across the entire trap frame \cite{cryo}, while glassy carbon contracts about an order of magnitude less \cite{htw}. Therefore, to limit the amount of sag in the glassy carbon rods, they are secured in a floating collar that uses a stainless steel spring to provide about 2 pounds of tension (Fig. \ref{fig:bpt}c). This decouples the tension applied to the rods from the thermal contraction of supporting frame.

To design the DC electrodes that provide axial confinement, SIMION was used to model thin planar electrodes at a larger radius than the rods. It was found that a stack of three electrodes, with RF+DC on the outer two and DC on the middle, was needed, as shown in Fig. \ref{fig:mkivview2}. This design was then optimized for the length of each electrode region (labeled as shown in Fig. \ref{fig:mkivview1}) to ensure a sufficiently large axial potential gradient in the center of the trap. The minimum length of the center region (``C") was limited by scattering and the allowable electrode region (previously described). The optimal center region length was determined to be 35 mm with an interior DC electrode (``B" and ``D") length of 40 mm. The interior electrodes ``B" and ``D" are made of graphite, a much less expensive option than glassy carbon, while the exterior electrodes ``A" and ``E," which are further away from the ion cloud, are made of stainless steel, as they have negligible impact on scattering. Graphite is a porous material and has been investigated for use in UHV environments over the decades but its use has been limited due to concerns over its purity, porosity, and contribution to out-gassing \cite{Beitel1971TheUO,Accettura_2019}. However, graphite also comes in a wide variety of grades with various porosities and purities and is an inexpensive material, even at high purities and relatively low porosities. Of significant advantage to us is that graphite is a conductor and that it can be easily machined to very high precision with wire electrical discharge machining. Tests at cryogenic temperatures indicated that the out-gassing rate of graphite grade G535 \cite{mwi} was UHV-compatible for use with the BPT Mk IV.

 The BPT Mk IV during assembly is shown in Fig. \ref{fig:bpt}. All elements are constructed from UHV compatible materials and were thoroughly cleaned before assembly. From \textsc{Geant4} simulations of the final design, 5.3\% of triple coincidence events came from scattered $\beta$, a roughly 4$\times$ reduction from the previous trap. 

\subsection{RF resonator}

\begin{figure}
\centering
\includegraphics[width=0.75\linewidth]{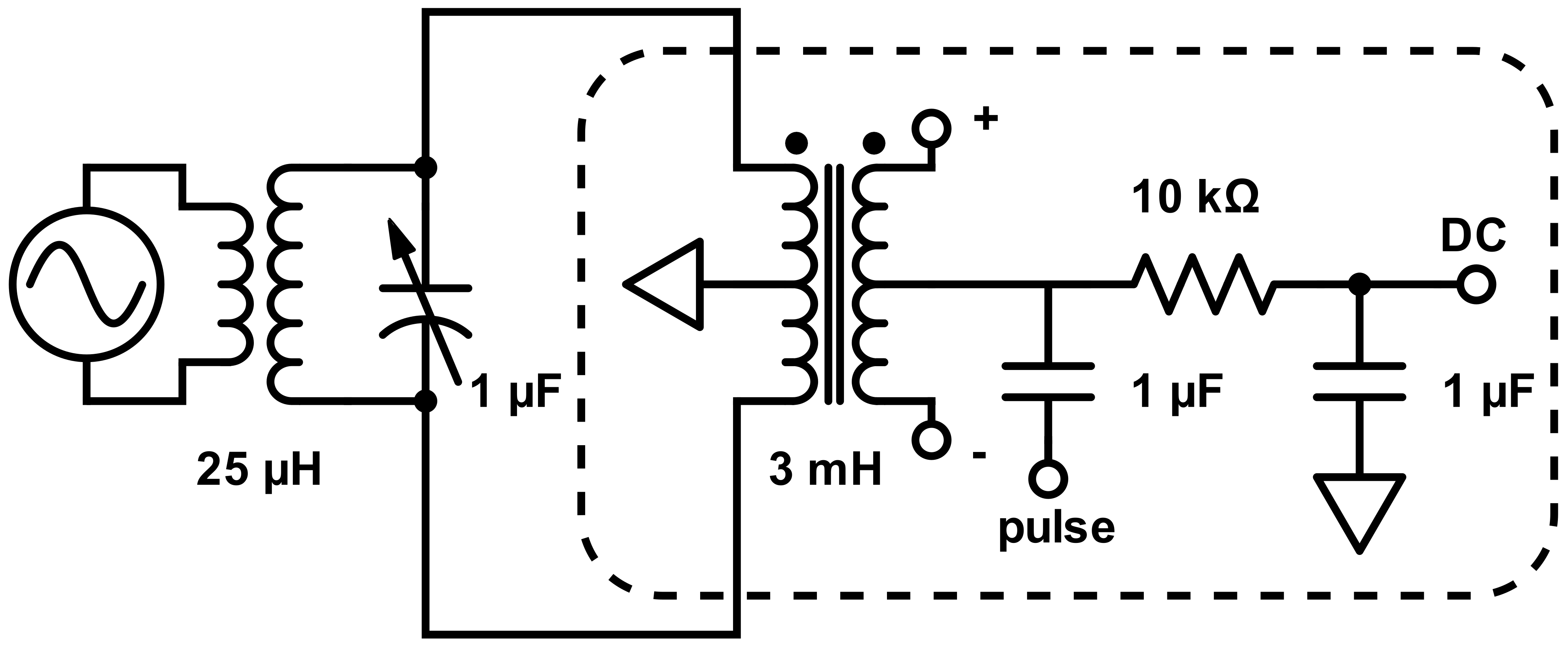}
\caption{Circuit diagram of the resonator for the BPT Mk IV. See text for details.}
\label{fig:resonator}
 \end{figure}

A new RF resonator circuit was required for the BPT Mk IV due to the increased number of electrodes, additional electrode regions, and the different electrical properties of the trap. The circuit design is shown in Fig. \ref{fig:resonator}. A primary $LC$ resonator circuit is shown on the left of the diagram, which uses a homemade air core transformer and an air variable capacitor. Five of the modules, indicated by dashed lines, are attached in parallel to the primary resonator circuit to create the independent circuits for each electrode pair. An in-house constructed common mode choke acting as a 1:1 transformer is used to provide a different DC offset to the electrode pairs while allowing each pair to have identical RF amplitudes and phases. The chokes are made using an N49 ferrite \cite{n49}, which has relatively low losses around the 1 MHz frequency range of interest.  Voltages for each electrode pair are taken from the ``+" and ``-" labeled sources, which have opposite RF phases but identical DC offsets. The DC input has a simple filter to attenuate any RF feedback to the power supply. A pulser is capacitively coupled to provide a fast voltage switch of up to a few hundred volts with $\sim 10$ ns rise time; these pulses are used to lower the electrode voltages during capture and ejection of the ions. 

The inductor on the primary resonator circuit is relatively low-valued due to a high trap capacitance of roughly $\sim600$ pF. This is likely due to the fact that thin Kapton wires with a coaxial ground shield were used to deliver the voltages to the trap in an attempt to reduce pickup inside of the chamber; these have a stated capacitance of 137 pF/m \cite{kapton}. To allow the circuit to achieve higher frequencies, an inductor can be added in parallel with the primary resonator circuit with a switch (not shown in Fig. \ref{fig:resonator}). The resonator circuit is tuned to the appropriate frequency with the variable capacitor. The resonator is driven by an RF amplifier (model T\&C ULTRA 2020), in turn driven by an SRS DS345 function generator \cite{srsds345} providing a sine-wave of the desired frequency. 

The quality factor of the resonator circuit is fairly low, with $Q\sim10$. Measurements of the RF voltage and phase on each electrode pair are difficult to perform accurately since attaching e.g. an oscilloscope to the electrodes slightly changes the properties of the circuit. With this in mind, the electrodes have amplitudes within $\lesssim 1$\% of each other and no phase mismatch was noticeable.

\section{Commissioning}

Offline testing of the BPT Mk IV was performed using argon and nitrogen gases. Ions were produced by introducing these gases into the upstream ion delivery system and ionizing them with a cold cathode gauge to produce $^{40}$Ar$^{+}$ and {$^{14}$N$_2$}$^+$. Trapping voltages and frequencies were found to be comparable to simulations. The BPT Mk IV was commissioned with \Li during a month-long data campaign at ATLAS. The \Li was produced through a $^7$Li($d,p$)$^8$Li reaction on a cryogenic deuterium gas target; details of the production and transport of the ions are available in Refs. \cite{Burkey_2019, Sternberg_2013}. After tuning the entire system for roughly a week, the trap voltage settings were finalized and held constant during the rest of the data collection. Data was collected for a total of 377 hours, resulting in around $2.6 \times 10^6$ triple coincidence events. This is a similar event rate as in the previous \Li campaign \cite{Burkey_Savard_Gallant_Scielzo_Clark_Hirsh_Varriano_Sargsyan_Launey_Brodeur_etal._2022}, which collected $2.0 \times 10^6$ triple coincidence events over a two week period. The typical trap population was estimated to be a few hundred \Li ions during each 12 s measurement cycle. The trapping lifetime was not measured directly but can be estimated from the trap population build-up during a measurement cycle using a model developed for the BPT \cite{louisthesis}. From a fit to this model, the trap lifetime was estimated to be 83 $\pm$ 29 seconds. This is not a precise determination but indicates that the trap lifetime is several times longer than each measurement cycle, long enough that it is not a concern for the experiment.

\begin{table}
\begin{center}
\begin{tabular}{c c c }
   electrode & DC (V) &pulse (V)\\
   \hline
 ``A" & +65  & -91\\
``B"  & -15  & \\
 ``C" (rods)  & -35  & \\
``D"  & -15  & -91\\
``E" & +66  & -100\\
\hline
RF & \multicolumn{2}{c}{1400 kHz, 300 V$_{\textrm{pp}}$ }\\
\hline
\end{tabular}
\caption{Voltage settings for the BPT Mk IV during the commissioning \Li run. Electrodes are labeled as in Fig. \ref{fig:trap}b. Electrode ``A" is pulsed during capture; electrodes ``D" and ``E" are pulsed during ejection. The resulting axial DC potential is shown in Fig. \ref{fig:potential}. }
\label{tab:bpt}
\end{center}
\end{table}

\begin{figure}
\centering
\includegraphics[width=0.75\linewidth]{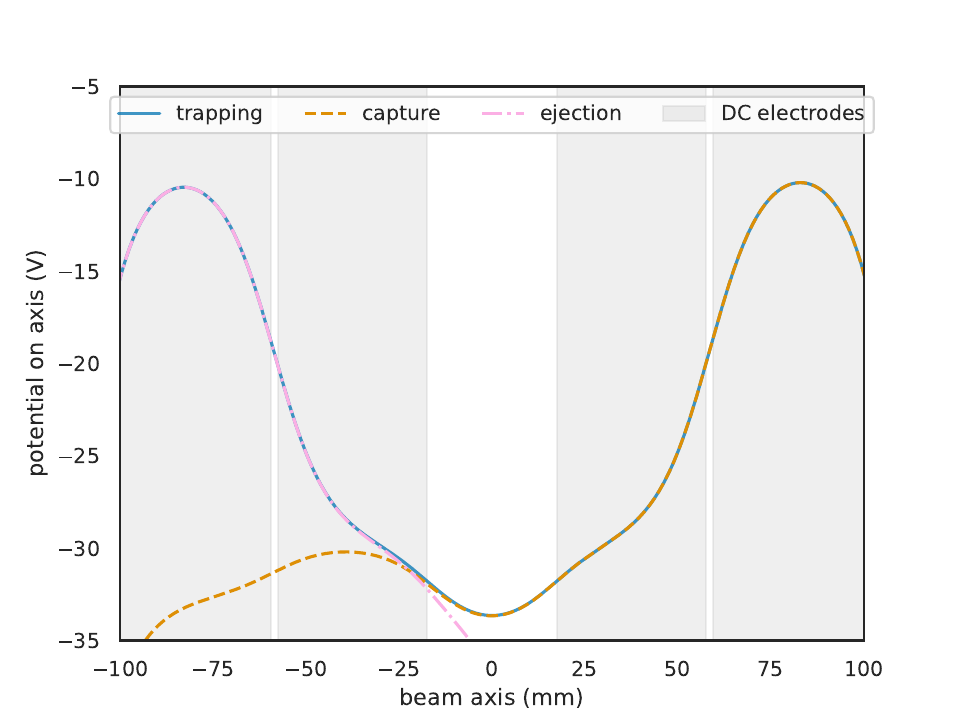}
\caption{Axial DC potential of the BPT Mk IV during the commissioning \Li experiment, as determined from SIMION using the applied voltages from the experiment. Applied voltages are shown in Table \ref{tab:bpt}. The potential during trapping (blue), during a capture pulse (orange) and during ejection (pink) are shown. The widths of the DC electrodes are shown as grey bands. The center region has no DC electrode, with the potential supplied directly onto the rods. }
\label{fig:potential}
 \end{figure}

A summary of the final trap settings is given in Table \ref{tab:bpt}, with the resulting axial potential shown in Fig. \ref{fig:potential}, as determined from SIMION using the applied voltages from the experiment. From these trap settings, the radial pseudo-potential has an effective field gradient of 0.042 V/mm$^2$ near the trap center with Mathieu parameter $q=0.32$, similar to that of the previous BPT \cite{Burkey_2019}. Since the glassy carbon rods are closer to the ions than the DC electrodes, there is a voltage-screening effect, necessitating a higher applied DC voltage on the electrodes to form a trapping potential on the axis. This also means that the capture and ejection pulses applied to the DC electrodes must be relatively high (roughly -100 V).  The helium buffer gas pressure was $\sim 1.2 \times 10^{-5}$ mbar as read by a cold cathode gauge in the trap vacuum chamber. The capture and trapping efficiency was estimated to be roughly $\sim25$\%; the efficiency could likely be improved through additional tuning.



\begin{figure}
     \centering
     \begin{subfigure}[b]{0.49\linewidth}
         \centering
         \includegraphics[width=\linewidth]{{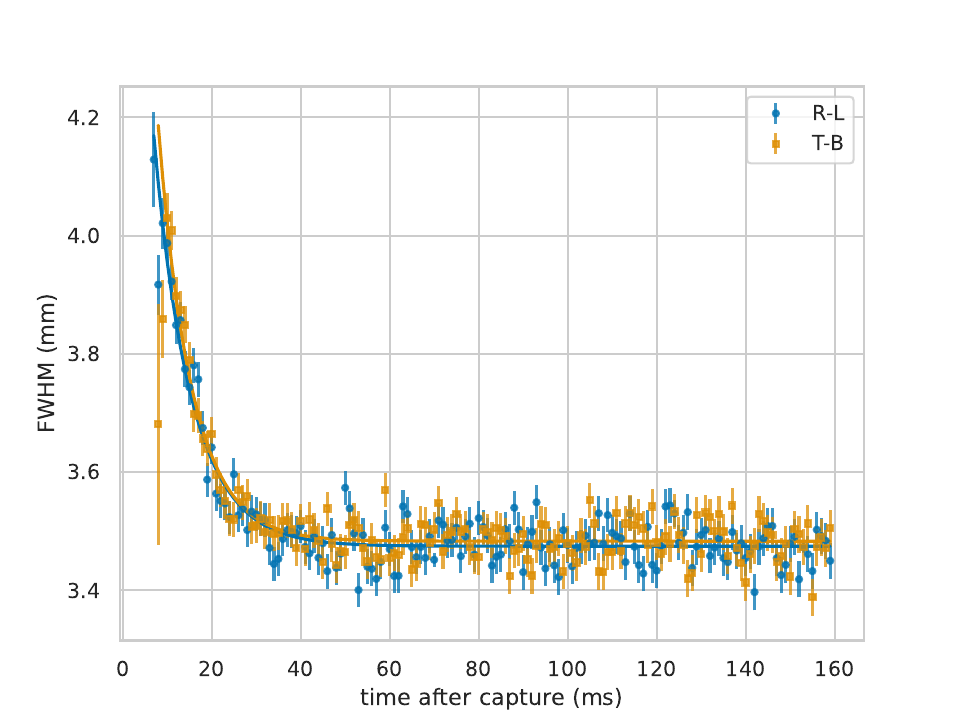}}
         \caption{axial cooling}
     \end{subfigure}
     \hfill
     \begin{subfigure}[b]{0.49\linewidth}
         \centering
         \includegraphics[width=\linewidth]{{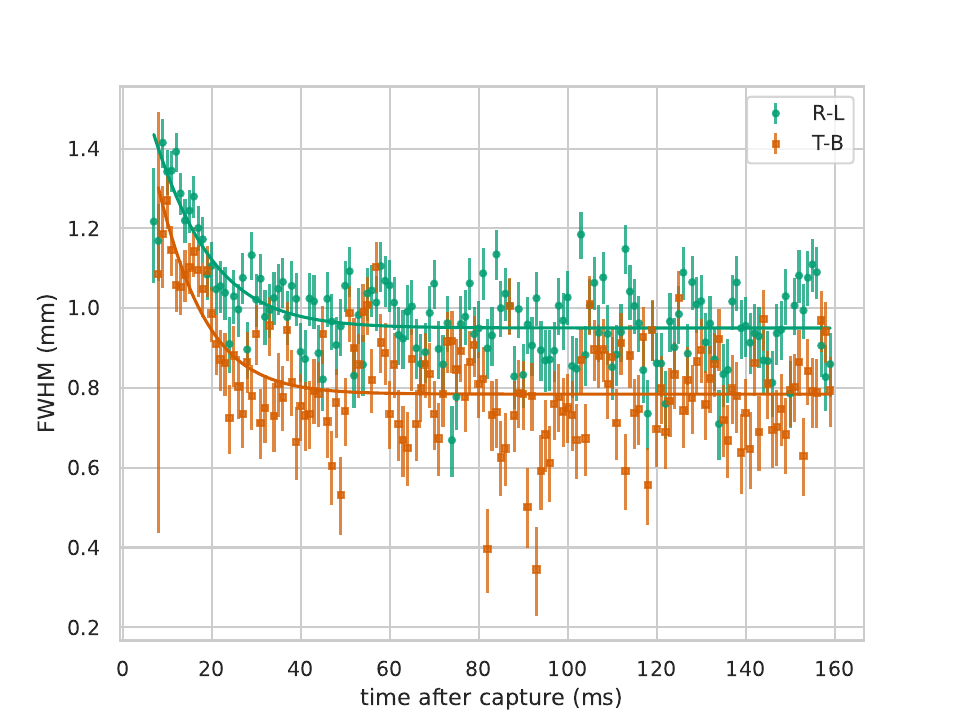}}
          \caption{radial cooling}
     \end{subfigure}
\caption{ \Li ion cloud cooling as a function of cycle time with (a) showing axial cooling and (b) showing radial cooling. The first few ms of data are unusable as noted in the text. The two different DSSDs pairs are shown separately in each plot (right-left, R-L, and top-bottom, T-B), with points and error bars indicating the fitted FWHM and its uncertainty at each 1 ms time slice. The solid lines show a fit to the data using an exponential decay plus a constant term. There is good agreement between the two pairs, and discrepancies may be due to non-functioning strips, which have not been accounted for.}
 \label{fig:cloud}
 \end{figure}

Due to the back-to-back $\alpha$ particles emitted from the \Li decay, the ion cloud may be directly self-imaged \cite{Scielzo_2012}. At each time slice, the spatial distribution of back-to-back $\alpha$ pairs on the DSSD surfaces is fit with a two-dimensional Gaussian distribution. The conversion between the observed distribution of $\alpha$ pairs and the spatial extent of the ion cloud is performed through a comparison to simulations of different ion cloud sizes. These simulations should be taken as an estimate only, with a precise determination to be performed in the course of the final data analysis. The ion cloud cooling may be observed during the trapping cycle, as shown in Fig. \ref{fig:cloud}. In this figure, the FWHM of the ion cloud and its uncertainty at every time slice is show for each of the detector pairs. A fit to an exponential decay plus a constant term is performed, resulting in the solid line that is intended primarily to guide the eye. The cooling time of about 30 ms is similar to that observed in previous BPT experiments \cite{Scielzo_2012}. Data from the first $6-8$ ms of each trapping cycle are unusable due to noise caused by the electrode pulsing. It is not well understood why this time period of unusable data is so long compared to the pulse length of a few microseconds. This is not a problem for the final data analysis, as this is during the initial ion cooling period of roughly $30$ ms and therefore would be discarded anyway.

A precise determination of the ion cloud size will be performed in the course of subsequent data analysis to determine $|C_T / C_A|^2$. From this initial estimate, however, the axial FWHM is close to 3.5 mm and the radial FWHM is about 0.9 mm. The expected size of the ion cloud can be estimated from the Maxwell-Boltzmann distribution at 80 K using the field gradient determined by SIMION from the applied voltages. The radial cloud size is about 20\% smaller than predicted, and the axial cloud size is about 13\% larger, but this is fairly good agreement. This final ion cloud size (accounting for the extent of 99.7\% of the ions) is larger than the design assumption, but a built-in tolerance means that this ion cloud size still has a full view of the detectors and the electrodes do not block decay products.

 \section{Conclusions}
 
 The BPT Mk IV has been successfully designed, constructed, and commissioned with a \Li data campaign, meeting design expectations. A new trap design incorporating rod electrodes made of glassy carbon and planar electrodes made of graphite has been demonstrated to operate under UHV conditions. Our simulations indicate that $\beta$ scattering, a key source of experimental systematic uncertainty, has been reduced by a factor of 4 from 21.1\% to 5.3\% of triple coincidence events. The commissioning experiment with \Li recorded over $2.6 \times 10^6$ triple coincidences, which will allow for $|C_T / C_A|^2$ to be more precisely measured than the latest BPT result and may allow for the recoil-order parameters to be experimentally determined, reducing the overall systematic uncertainty of the experiment.
 
  \section{Acknowledgments}

  This work was carried out under the auspices of the U.S. Department of Energy, by Argonne National Laboratory under Contract No. DE-AC02-06CH11357 and Lawrence Livermore National Laboratory under Contract No. DE-AC52-07NA27344. Funding by NSERC (Canada) under Contract SAPPJ-2018-0028 and the U. S. National Science Foundation under grant PHY-2011890 are acknowledged. L. V. was supported by a National Science Foundation Graduate Research Fellowship under Grant No. DGE-1746045. This research used resources of Argonne National Laboratory’s ATLAS facility, which is a DOE Office of Science User Facility.

\appendix

\bibliographystyle{elsarticle-num-names} 
\bibliography{cas-refs}





\end{document}